\documentclass[article]{aa}

\usepackage{txfonts}
\usepackage{graphicx}
\usepackage{natbib}
\bibpunct{(}{)}{;}{a}{}{,}

\begin{document}

\titlerunning{A chromospheric flare}
\authorrunning{P.H. Keys et~al.}

\title{Chromospheric velocities of a C-class flare}

\author{Peter H. Keys\inst{1}, David B. Jess\inst{1}, Mihalis Mathioudakis\inst{1} \and Francis P. Keenan\inst{1}}

\offprints{P.H. Keys, \email{pkeys02@qub.ac.uk}}

\institute{Astrophysics Research Centre, School of Mathematics and Physics, Queen's University, Belfast, BT7 1NN, Northern Ireland, U.K.}

\date{Received 15th December 2010 / Accepted 10th March 2011}

\abstract 
{} 
{We use high spatial and temporal resolution observations from the Swedish Solar Telescope to study the chromospheric velocities of a C-class flare originating from active region NOAA 10969.}
{A time-distance analysis is employed to estimate directional velocity components in H$\alpha$ and Ca~{\sc{II}}~K image sequences. Also, imaging spectroscopy has allowed us 
to determine flare-induced line-of-sight velocities. A wavelet analysis is used to analyse the periodic nature of associated flare bursts.} 
{Time-distance analysis reveals velocities as high as 64~kms$^{-1}$ along the flare ribbon and 15~kms$^{-1}$ perpendicular to it. The velocities are very 
similar in both the H$\alpha$ and Ca~{\sc{II}}~K time series. Line-of-sight H$\alpha$ velocities are red-shifted with values up to 17~kms$^{-1}$. The high spatial and 
temporal resolution of the observations have allowed us to detect velocities significantly higher than those found in earlier studies. Flare bursts with a periodicity 
of $\approx$60~s are also detected. These bursts are similar to the quasi-periodic oscillations observed at hard X-ray and radio wavelength data.}
{Some of the highest velocities detected in the solar atmosphere are presented. Line-of-sight velocity maps show considerable mixing of both the magnitude and direction of velocities along the flare path. 
A change in direction of the velocities at the flare kernel has also been detected which may be a signature of chromospheric evaporation.}

\maketitle 

\section{Introduction}
\label{intro}
A flare is the result of magnetic field reconfiguration/annihilation, a process usually occurring in the corona \citep{Park63}. 
Solar flares are highly energetic events with complex structures and properties, which transfer 
energy between various regions in the solar atmosphere. White-light flares (WLFs) exhibit strong emission in the white-light continuum, and are 
usually present in the most energetic of events, such as X- and M-class flares 
\citep{Nei83}. However, with the increasing use of space-based telescopes and ground-based observatories with 
high spatial resolutions, flares as low as C-class have been shown to demonstrate optical continuum enhancements 
\citep{Matt03, Hud06}. \citet{DBJ08} have recently suggested that {\it{all}} flares exhibit a degree of continuum enhancement, 
but that most observations will not have sufficient spatial or temporal resolution to capture this phenomena. 
The processes which ultimately lead to WL enhancements are still under review \citep{Fle07}, particularly in the regime of 
small flare classifications. 

\citet{Asch07} have suggested that the principal source of coronal heating during flare events may be in the transition region and the upper 
chromosphere. 
After the reconnection of magnetic field lines has occurred, the lower solar atmosphere is heated by either 
chromospheric evaporation or radiative back-warming. It is believed that the process which governs the heating of the lower 
solar atmosphere is determined by the magnitude of the flaring event. There have been flare models 
which explain two types of chromospheric evaporation: gentle evaporation \citep{Ant78, Mar89} 
and explosive evaporation \citep{Fis85, Mil06a, Mil06b}. Gentle evaporation refers to a situation 
where the chromosphere is heated directly by non-thermal electrons, or heated indirectly by thermal 
conduction processes throughout the solar atmosphere. The chromospheric plasma will then lose energy 
through radiative processes and low-velocity hydrodynamic expansion. With explosive evaporation, 
the chromosphere is unable to radiate energy at a sufficiently high rate, resulting in 
the plasma expanding into the flaring loops at high velocities. 

The other proposed heating mechanism is radiative back-warming \citep{Mac89}. Under this regime, electron 
beams bombard the chromosphere after magnetic reconnection occurs in the corona. However, 
most of the electrons will not have sufficient energy to pass through the dense chromospheric 
plasma, and so will not reach the photosphere. On the other hand, physical conditions linking the 
chromosphere and photosphere will promote the movement of energy towards the photosphere; 
hence the term back-warming. This process is not as impulsive as chromospheric evaporation, 
and may help explain the origin of WL emission seen in low energy events.

The present paper uses high temporal, spatial and spectral resolution observations to study the chromospheric velocities of a C-class flare. In \S~\ref{Obs} 
we discuss the observations and image reconstruction techniques employed, while \S~\ref{R&L} describes the analysis methods employed on the dataset and 
presents our results. Concluding remarks are given in \S~\ref{Conc}. 

\section{Observations}
\label{Obs}
Data presented here are part of an observing sequence obtained on 2007 August 24, with the Swedish Solar Telescope (SST) 
on the island of La~Palma. The optical setup allowed us to image a $68\arcsec \times 68\arcsec$ region surrounding active 
region NOAA~10969, complete with solar rotation tracking. This active region was located at heliocentric 
co-ordinates ($-516\arcsec$,$-179\arcsec$), or S05E33 in the solar NS-EW co-ordinate system. The Solar Optical Universal 
Polarimeter (SOUP) was implemented to provide two-dimensional spectral information across the H$\alpha$ line 
profile centred at $6562.8$~{\AA}, which was sampled using 7 wavelength positions followed by a Gaussian fit to the line profile. 
Doppler velocities were established using the methods detailed in \citet{Sue95} and a Doppler map, $Dopp = (C_{b} - C_{r}) / (C_{b} + C_{r} + 2)$, was constructed. 
In this equation, $C_{b}$ and $C_{r}$ are contrast images obtained using the relation $C = (I-I_{a})/I_{a}$ where $I_{a}$ 
corresponds to the average intensity over the entire dataset, while the subscripts $b$ and $r$ correspond to 
the wavelengths at H$\alpha$-core $-700$~m{\AA} and H$\alpha$-core $+700$~m{\AA}, respectively. A second camera provided simultaneous 
wideband images centered with a FWHM of 8~\AA\ centered in H$\alpha$ \citep{Lan08}.

In addition, a series of Ca~{\sc{ii}} interference filters were used to provide 
high-cadence imaging in this portion of the optical spectrum. A filter centred at $3953.7$~{\AA} with a band width 
of $10$~{\AA} was employed to acquire data in the Ca~{\sc{ii}}~K/H continuum whilst a $1.5$~{\AA} filter band centred at $3934.2$~{\AA} was used to acquire data in Ca~{\sc{ii}}~K.

The observations employed in the present analysis consist of 39000 images at each wavelength, taken with a 0.12~s cadence, 
providing just over one hour of uninterrupted data. Our H$\alpha$ images have a sampling of $0.068\arcsec$ per 
pixel, to match the telescope's diffraction-limited resolution to that of the $1024 \times 1024$~pixel$^{2}$ CCD. Images of 
the Ca~{\sc{ii}}~K core were captured using a $2048 \times 2048$~pixel$^{2}$ CCD with a sampling of $0.034\arcsec$ per 
pixel. Although the Ca~{\sc{ii}}~K camera was oversampled, this was deemed necessary to keep the dimensions of the 
field-of-view the same for both cameras. A high order adaptive optics system that corrects for approximately 35  Karhunen-Loeve 
modes was utilized throughout the data acquisition \citep{Sch03}. The acquisition time for this observing sequence was early in the 
morning, and seeing conditions were excellent with minimal variation throughout the time series. During the observing sequence a 
C2.0 flare was observed, originating from NOAA~10969 at 07:49~UT.  

Multi-Object Multi-Frame Blind Deconvolution \citep[MOMFBD;][]{van05} image restoration was implemented to 
remove small-scale atmospheric distortions present in both data sets. Sets of 80 exposures were included in the restorations, producing 
a new effective cadence of 9~s for the H$\alpha$ observations. The Ca~{\sc{ii}}~K core data incorporated 20 exposures in each restoration, 
producing a new effective cadence of 2.5~s. All reconstructed images were subjected to a Fourier co-aligning routine, in which 
cross-correlation and squared mean absolute deviation techniques are utilized to provide sub-pixel co-alignment accuracy. The full field-of-view with 
expanded versions of the flare is shown in Figure~\ref{Fig1}. Movies of the event as seen in the H$\alpha$ wideband and Ca~{\sc{ii}}~K core 
spectral lines have been included as an on-line resource material to emphasise the high temporal and spatial resolution of this dynamic event.

\section{Results and discussion}
\label{R&L}
\subsection{Flare Velocities}
\label{Vel}
The horizontal velocity components in the H$\alpha$ wideband and Ca~{\sc{ii}}~K core images are investigated using time-distance analysis. 
We place a one-dimensional slice through each of the data sets to generate intensity maps as a function of distance and time.

\begin{figure*}[h!]
\centering
   \includegraphics[width=12.0cm]{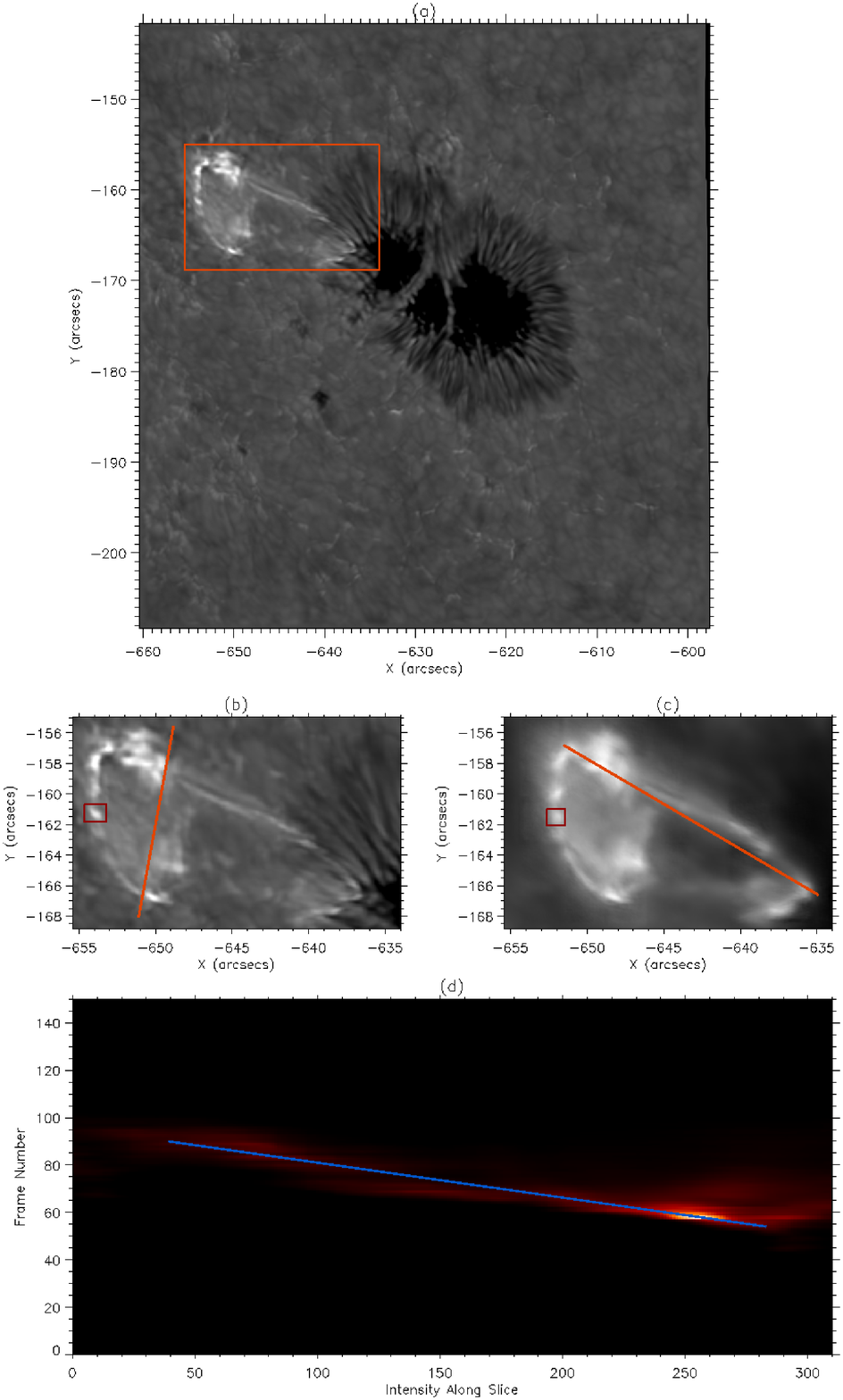}
     \caption{(a) A snapshot of the full 68" x 68" field-of-view captured through the H$\alpha$ wideband filter. 
     An expanded view of the flare region in H$\alpha$ (b)  and Ca~{\sc{ii}}~K core (c) are also shown with sample positions 
     for the time-distance slice method used in \S~\ref{Vel} and \S~\ref{Oscillations}. A box is included in both image (b) 
     and (c) to highlight one of the brightenings used to determine the horizontal bulk plasma velocities summarised in Table~\ref{table1}. 
     A sample time-distance plot using the slice position indicated in (c) is shown also (d). The blue line shows the gradient used to establish the 
     velocity.
     }
     \label{Fig1}
\end{figure*}

The slice was initially positioned at the flare kernel \citep{DBJ08}, and 
extended along the flare ribbon, allowing accurate velocity measurements as a function of distance from this 
reference point (Fig.~1c). Errors are evaluated by fitting the data with a least squares fit. The temporal 
evolution of the chromospheric velocities for Ca~{\sc{ii}}~K and H$\alpha$, determined using the slice method, are shown in Figure 2. 
Velocities peak at a value of $64 \pm 11$~kms$^{-1}$ in both wavelengths, and then these are reached $\approx$90~s after 
the Ca~{\sc{ii}}~K intensity peak. To minimize the effects of small-scale fluctuations along the flare ribbon, we have also 
determined the velocities using a  wider slice of 15 pixels, revealing identical results. The velocities found are within the 
error bars of those determined with a narrow slice, suggesting the plasma flows uniformly along a direct path towards the flare kernel. 
Bulk plasma velocities are also calculated by tracking individual flare brightenings in the H$\alpha$ data and the Ca~{\sc{ii}}~K core images.  
Table~\ref{table1} summarizes the velocities of three brightenings, with one such brightening enclosed in a box in Figure~\ref{Fig1}b and \ref{Fig1}c.
We find these bulk motions do not follow the general movement of the flare ribbon, but instead they move independently across the top of the flare structure.

Velocity components perpendicular to the flare ribbon are also estimated using a similar approach. 
This form of analysis is employed on the Ca~{\sc{ii}}~K datacube, as its superior temporal resolution allows small-scale 
velocities to be disentangled with a high degree of precision. An average velocity of $15.5 \pm 0.6$~kms$^{-1}$ is 
found perpendicular to the flare ribbon, with velocities directed towards the flare ribbon.

\begin{table}[h!]
\caption{Velocities of Bright Points}           
\label{table1}      
\centering                                    
\begin{tabular}{l c c c}          
\hline\hline                        
Spectral &  & Bright Point Velocity (kms$^{-1}$) & \\
     Line & 1 & 2 & 3 \\				 
\hline                                   
    Ca~K & $44 \pm 10$ & $45 \pm 10$ & $37 \pm 10$\\     
    H$\alpha$ & $49 \pm 5$ & $55 \pm 5$ & $47 \pm 5$\\
\hline                                            
\end{tabular}
\end{table}

The spectroscopic information provided by the SOUP instrument allows us to derive the line-of-sight (LOS) Doppler velocity components of the flare ribbon with an 
effective cadence of 63~s \citep{DBJ10}. In Figure 3 we show the velocity maps derived from the H$\alpha$ scans. The average LOS velocity in the immediate 
viscinity of flare-induced brightenings was found to be 
$14.83 \pm 0.29$~kms$^{-1}$, peaking at 22.5~kms$^{-1}$. 
Further examination of the Doppler maps reveals a shift in velocity direction at the location of the flare kernel between the first and second 
scans. The first frame exhibits 
a red-shifted velocity of $6.06\pm0.28$~kms$^{-1}$, while the subsequent frame shows a blue-shifted velocity of 
$14.8\pm0.3$~kms$^{-1}$ at the same position. It is possible that this shift from a red-shifted to a blue-shifted 
velocity is a signature of a chromospheric evaporation process, whereby accelerated electrons/plasma strike the lower solar atmosphere, resulting in 
material being ejected into the flaring loops. Figure~\ref{Fig3} also highlights considerable mixing of both the magnitude and 
direction of flare-induced velocities over the entire flare ribbon.

Velocities of $54.5 \pm 2.3$~kms$^{-1}$ and $52.8 \pm 2.5$~kms$^{-1}$ for the H$\alpha$ and Ca~{\sc{ii}}~K core datacubes, respectively, show that 
velocities along the flare ribbon are very similar in these chromospheric lines. Utilizing a wider slit, followed by coarse spatial averaging, produced velocities of 
$58.1 \pm 2.2$~kms$^{-1}$ and $59.0 \pm 1.8$~kms$^{-1}$ for the H$\alpha$ wideband and Ca~{\sc{ii}}~K core images, respectively. 
Again, these values reiterate flare-induced velocities that are consistent along a direct path towards the flare kernel. We emphasize that, given the broad nature 
of the H$\alpha$ line profile, the velocities determined from the H$\alpha$ wideband are chromospheric 
 
\citet{Fal06} examined the Ca~{\sc{ii}}~K spectral line of a C1.0 flare and found LOS velocities of 10--20~kms$^{-1}$, 
with He~{\sc{i}} and O~{\sc{v}} observations indicating  
velocities closer to 40~kms$^{-1}$. During the same observing campaign, \citet{Ter06} report on a larger C2.3 class flare 
which occurred within the same active region, with He~{\sc{i}} and O~{\sc{v}} velocities of $\approx$20~kms$^{-1}$. The velocities 
presented here are substatially larger than previously published results for C-class flares.

Moving on to more energetic events, \citet{Mau09} observed transient features associated with X17 and 
X10 WLFs and derived photospheric and chromospheric 
velocities of 30--50 km$s^{-1}$. Furthermore, work by \citet{Mil06a} found X-class flare-induced downflows of $36 \pm 16$~kms$^{-1}$ at 
chromospheric temperatures. The values established 
for the presented Ca~{\sc{ii}}~K core data, in particular, show remarkable agreement with the chromospheric velocity results 
determined by \citet{Mau09}, albeit originating from a much lower energy event. 

We believe that the higher velocities established here for less-energetic events are a direct result of the superior spatial 
and temporal resolution offered by the SST, which can be clearly seen in the movies provided as an on-line resource. Indeed, the velocities uncovered here are consistent with current 
radiative hydrodynamic simulations of solar flares, where \citet{All05} 
determine the response of the solar atmosphere to a beam of non-thermal electrons, injected at the apex of a closed coronal loop. The 
results of their simulations for typical M- and X-class flares produce velocities 
of $\sim$40~kms$^{-1}$ in the lower chromosphere. Thus, the velocities presented here demonstrate a striking resemblence to current theoretical models of solar flares.

\subsection{Periodic variations}
\label{Oscillations}
Analysis of the lightcurve associated with the flare revealed 4 periodic bursts (Fig.~\ref{Fig4}) with an average periodicity of $\approx$60~s. 
The period was determined using wavelet analysis \citep{Tor98}. Bursts were found to exist at a distance of $\approx$30~Mm from the flare kernel, in a region encompassing 
the main loop structure of the flare (Fig.~\ref{Fig1}). A control test was performed over a quiet region of the data set using the same technique employed on the flare 
region. The outcome of this test confirmed that the intensity fluctuations were not the result of variations in seeing conditions.

Periodic flare brightnenings are often interpreted in terms of quasi-periodic pulsations (QPPs) \citep{Nak09}. QPP studies \citep{Nak10} have 
found periods of 40~s in gamma-ray, hard X-ray and microwave emissions. Therefore, it is possible 
that the 40--80~s oscillations found in optical emissions within this study are a result of flare-induced QPPs. Additionally, work by 
\citet{McA05} found periods between 40--80~s along the ribbon of a C9.6 class flare. Flare-induced acoustic waves in the overlying 
coronal loops were thought to be responsible for the periodic brightenings in this study. Although the periodic bursts presented here are 
observed away from the flare kernel, the fact that the observed periods agree with the study of 
\citet{McA05} suggests these oscillations extend beyond the boundaries of the magnetic flux tube and into the chromosphere.

\section{Conclusions}
\label{Conc}
We have utilized high spatial and temporal resolution imaging, and imaging spectroscopy of the lower solar atmosphere to derive 
velocities associated with a relatively low energy C-class flare. Velocities across the solar surface have values as high 64~kms$^{-1}$, while line-of-sight (LOS) velocities 
show both redshifts and blueshifts up to 17~kms$^{-1}$. These are some of the highest velocities detected in the lower solar atmosphere. Bulk plasma motions were 
observed to move across the flare ribbon at velocities as high as 55~kms$^{-1}$, seemingly independent of the general flare ribbon motion. A time-distance slice employed 
parallel to the flare ribbon found velocities as high as 64~kms$^{-1}$ for H$\alpha$ wideband data and as high as 59~kms$^{-1}$ for Ca~{\sc{ii}}~K core data. Varying spatial 
averages of the time-distance slice produced similar values, indicating a uniform motion on large spatial scales. 
Analysis of the LOS components of the flare found velocities as high as 17~kms$^{-1}$, with substantial mixing of both magnitude and direction 
of flare-induced velocities. Also, a velocity direction change at the flare kernel was observed in the LOS Doppler maps, which could be attributed to a form of chromospheric evaporation. 
Examination of the light curve associated with the flare event displays a periodic nature, whereby bursts are emitted at $\approx$60~s intervals. 
In addition, a comparison with more energetic flares indicate that flare-induced velocities are not necessarily 
directly related with the magnitude of the event. 

\begin{acknowledgements}
PHK thanks the Northern Ireland Department for Employment and Learning for a PhD studentship. 
DBJ thanks the Science and Technology Facilities Council for a Post-Doctoral Fellowship. FPK is 
grateful to AWE Aldermaston for the award of a William Penney Fellowship. The Swedish 1-m 
Solar Telescope is operated on the island of La Palma by the Institute for Solar Physics of 
the Royal Swedish Academy of Sciences in the Spanish Observatorio del Roque de los Muchachos 
of the Instituto de Astrof\'{i}sica de Canarias. These observations have been funded by the 
Optical Infrared Coordination network (OPTICON), a major international collaboration 
supported by the Research Infrastructures Programme of the European Commissions Sixth Framework 
Programme.
\end{acknowledgements}

\bibliographystyle{aa}

\clearpage

\clearpage

\begin{figure*}[h!]
\centering
   \resizebox{\hsize}{!}{\includegraphics{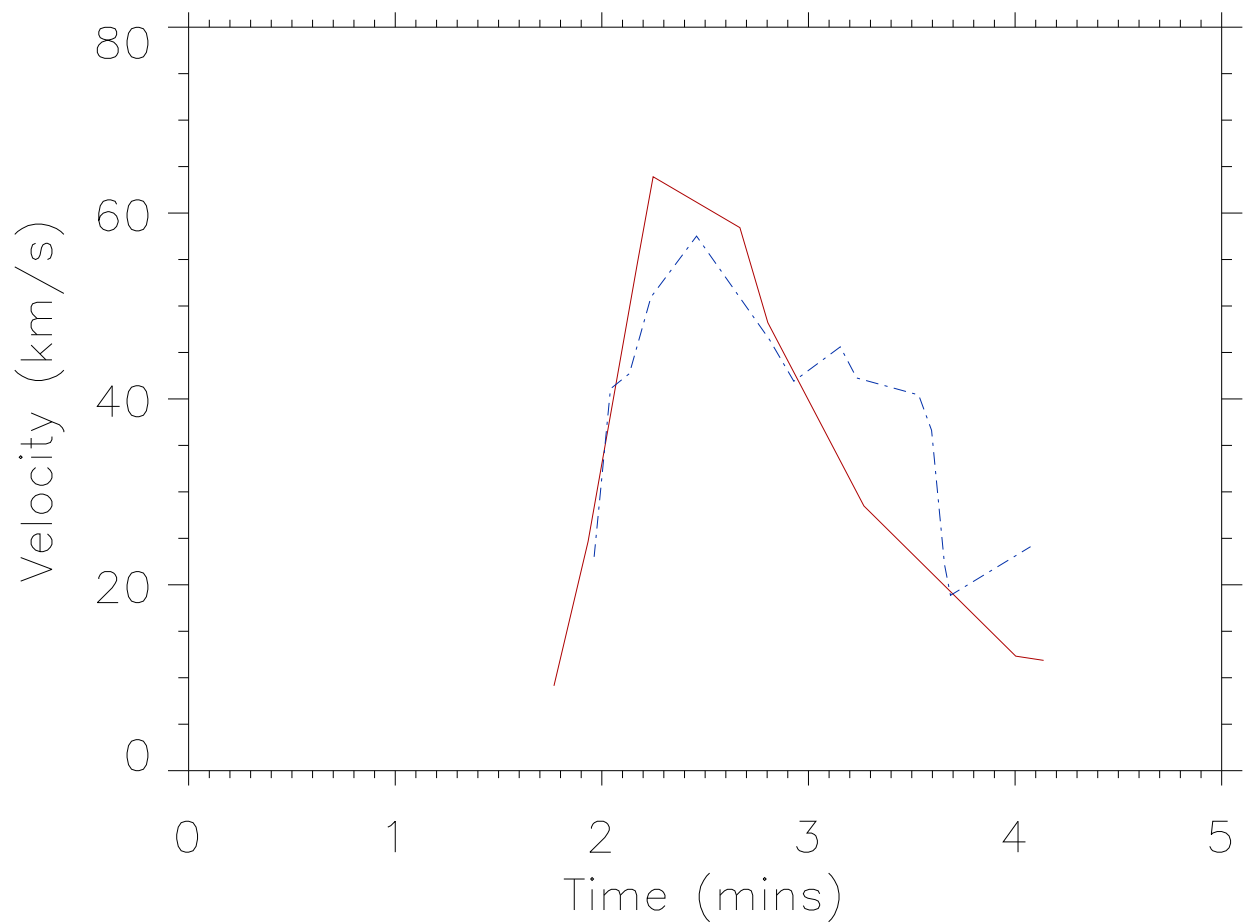}}
     \caption{The evolution of flare velocities in H$\alpha$ ({\it{solid red line}}) and Ca~{\sc{ii}}~K core ({\it{dot-dashed blue line}}) images, determined using a slice along the flare ribbon. 
     The data sequence begins at 07:50:34 UT. 
     }
     \label{Fig2}
\end{figure*}

\clearpage

\begin{figure*}[h!]
  \resizebox{\hsize}{!}{\includegraphics{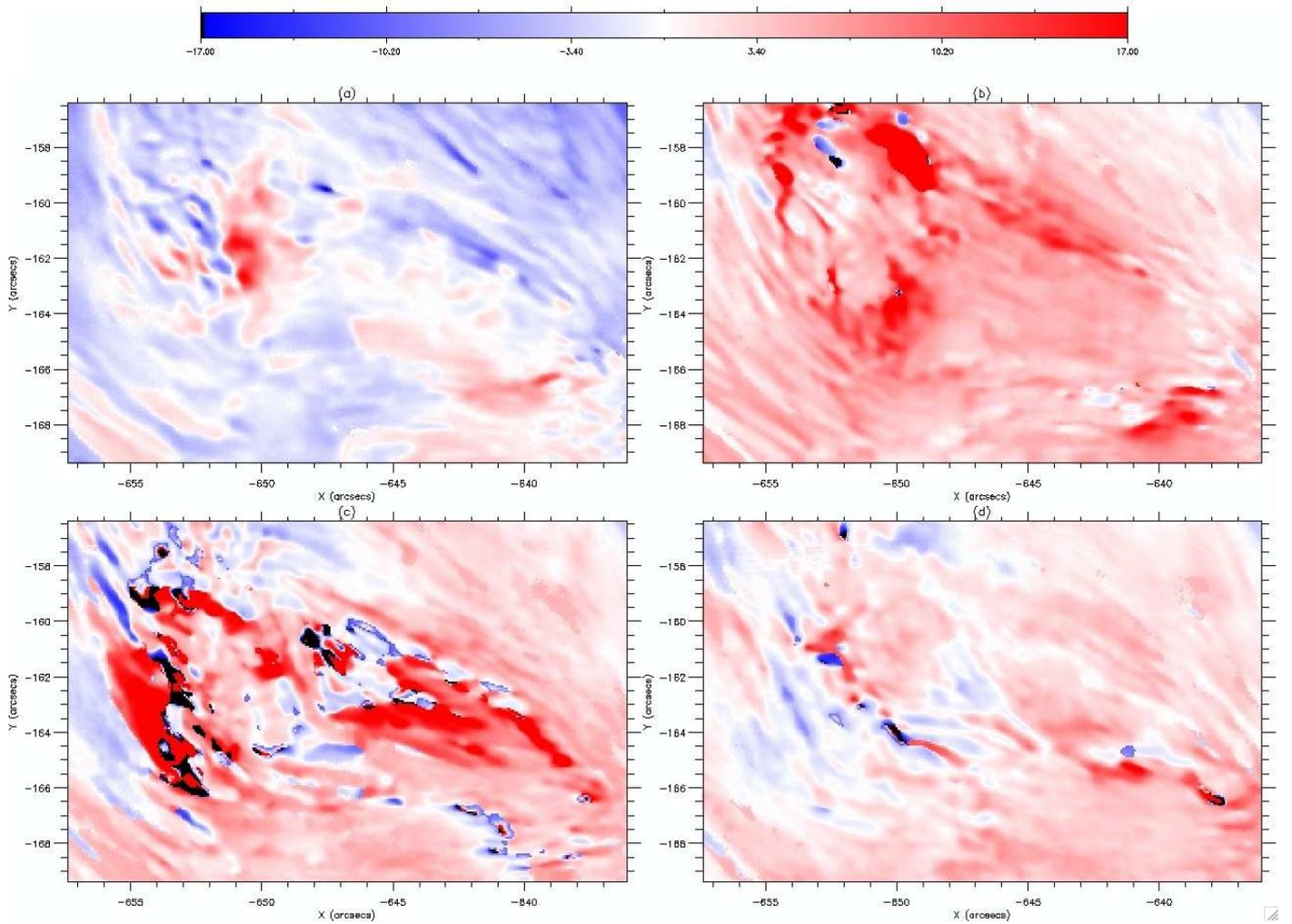}}
  \caption{Line-of-sight velocities of the flare region determined from the H$\alpha$ SOUP scans. (a), (b), (c) and (d) are velocity maps obtained at 
  07:50:40~UT, 07:51:23~UT, 07:52:26~UT and 07:53:29~UT respectively. The colour scale represents line-core Doppler shifts and is displayed in kms$^{-1}$.}
  \label{Fig3}
\end{figure*}

\clearpage

\begin{figure*}[h!]
  \resizebox{\hsize}{!}{\includegraphics{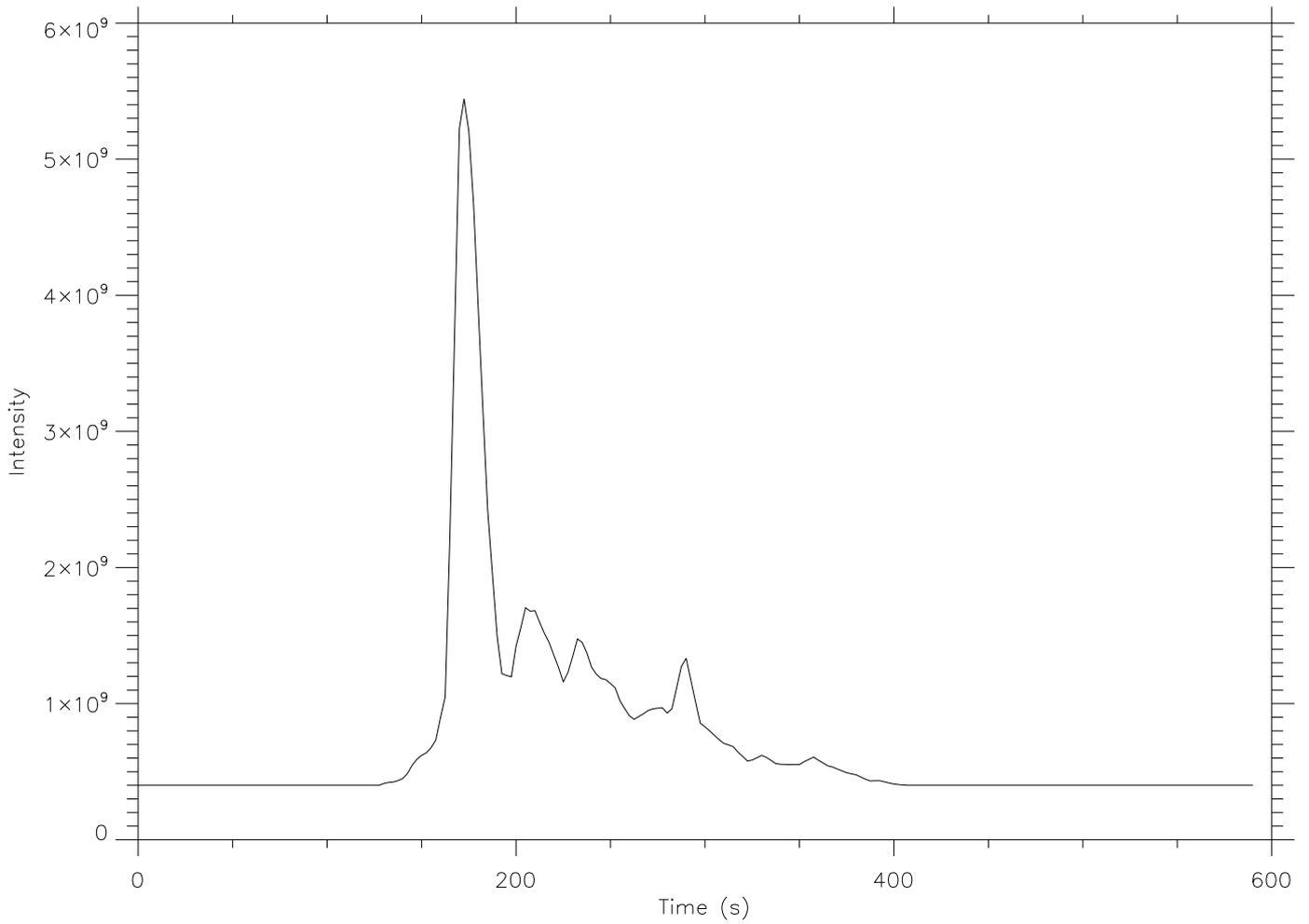}}
  \caption{ A sample light curve of the flaring event generated by methods discussed in \S~\ref{Oscillations}. This light curve clearly demonstrates the 
  periodic nature of the flare bursts. The average period of the flare bursts was $\approx$60~s, and all values were 
  found to be well above the 3$\sigma$ confidence level.}
  \label{Fig4}
\end{figure*}

\clearpage

\begin{figure*}[h!]
\centering
	\includegraphics[width=12.0cm]{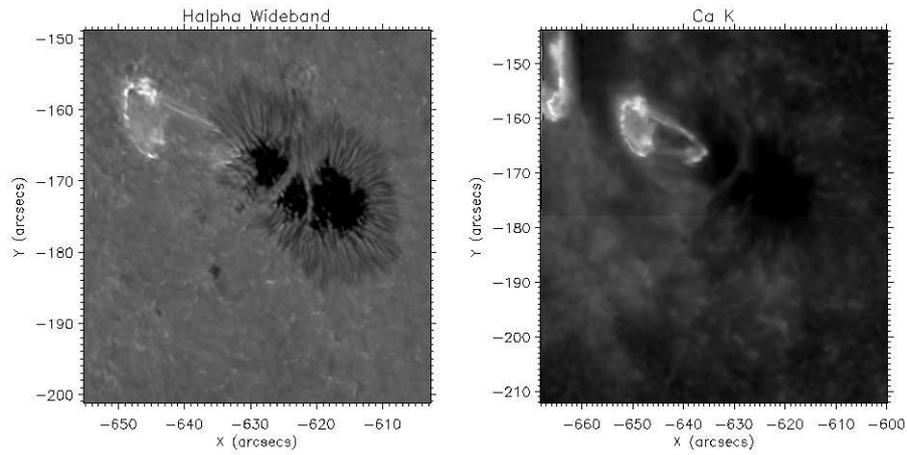}
	\caption{Two movies illustrating the temporal and spatial evolution of a C2.0 class flare observed on 2007 August 24 in active region NOAA 10969 with 
	the Swedish Solar Telescope. The data have been obtained with a wide band filter centered in H$\alpha$ (FWHM 8~{\AA}) and a filter centered in the Ca~{\sc{ii}}~K 
	core (FWHM 1.5~{\AA}). This flare has shown a white-light continuum previously reported by Jess et al. (2008). The first frame in the Ca~K movie was obtained at 
	07:50:30~UT and the corresponding frame in the H$\alpha$ wideband was acquired at 07:50:40~UT. Imaging spectroscopy of the event revealed high chromospheric 
	velocities and quasi-periodic oscillations similar to those detected in hard X-ray and radio wavelengths. The results and methodology are discussed in detail 
	in the text. 
	}
	\label{movie}
\end{figure*}

\end{document}